\documentclass[12pt]{article}

\begin{document}
\pagestyle{empty}
\begin{flushright}
\end{flushright}
\vspace*{3mm}
\begin{center}
{\Large \bf
Order $\alpha_s^2$ perturbative QCD corrections
to\\ the Gottfried sum rule}

\vspace{0.1cm}

{\bf A.L. Kataev$^{a}$ and G. Parente$^{b}$}\\
\vspace{0.1cm}
$^{a}$
Institute for Nuclear Research of the Academy of Sciences of
Russia,\\ 117312, Moscow, Russia\\
$^{b}$
Department of Particle Physics, University of Santiago de Compostela,\\
15706 Santiago de Compostela, Spain\\

\end{center}
\begin{center}
{\bf ABSTRACT}
\end{center}
\noindent
The order $\alpha_s^2$ perturbative QCD correction 
to the Gottfried sum rule is  obtained. The result is based on numerical calculation 
of the order $\alpha_s^2$ contribution  to the coefficient function 
and on the new  estimate of the three-loop  anomalous dimension term. 
The  correction found  is negative and  rather small. Therefore 
it  does not affect 
the necessity to introduce flavour-asymmetry between $\overline{u}$ and 
$\overline{d}$ antiquarks for the description of NMC result for the 
Gottfried sum rule.  
\vspace*{0.1cm}
\noindent
\\[3mm]
\vfill\eject

\setcounter{page}{1}
\pagestyle{plain}

\section{Introduction}

One of the still actively discussed problems of deep-inelastic scattering 
(DIS) is related to the consideration  of the Gottfried sum rule  
\cite{Gottfried:1967kk}, namely
\begin{eqnarray}
I_{GSR}(Q^2)&=&\int_0^1\bigg[F_2^{lp}(x,Q^2)-F_2^{ln}(x,Q^2)\bigg]\frac{dx}{x}
\\ \nonumber
&=&\int_0^1\bigg[\frac{1}{3}\bigg(u_v(x,Q^2)-d_v(x,Q^2)\bigg)+\frac{2}{3}
\bigg(\overline{u}(x,Q^2)-\overline{d}(x,Q^2)\bigg)\bigg]dx\\ \nonumber 
&=&
\frac{1}{3}+\frac{2}{3}\int_0^1
\bigg(\overline{u}(x,Q^2)-\overline{d}(x,Q^2)\bigg)dx~~~~.       
\end{eqnarray}
This sum rule is  
the {\it first} non-single (NS) moment of the difference of $F_2$
structure functions (SFs) of 
charged lepton-nucleon DIS which in general has the following 
definition: 
\begin{equation}
M_n^{NS}(Q^2)=\int_0^1x^{n-2}\bigg[F_2^{lp}(x,Q^2)-F_2^{ln}(x,Q^2)\bigg]dx
\end{equation}
An extensive discussion of the current studies of this sum rule 
was given in the review of Ref. \cite{Kumano:1997cy}. However, for the sake 
of completeness, we will remind the existing experimental situation,which 
is stimulating the continuation of the research of various subjects, 
related to the Gottfried sum rule.

In fact if the sea is flavour symmetric, namely $\overline{u}$=$\overline{d}$,
one should have 
\begin{equation}
\label{FS}
I_{GSR}=\frac{1}{3}~~~~.
\end{equation}
However, the most detailed analysis of muon-nucleon DIS data of NMC 
collaboration  
gives the following 
result \cite{Arneodo:1994ia}:
\begin{equation}
I_{GSR}^{exp}(Q^2=4~{\rm GeV}^2) = 0.235 \pm 0.026~~~. 
\end{equation}
It   clearly indicates the violation of  theoretical
expression of Eq.(\ref{FS}) and necessitates more detailed investigations 
of different 
effects, related to the Gottfried sum rule. In this Letter we 
reconsider the question of studying the contributions of $\alpha_s^2$, 
corrections to this sum rule previously raised in Ref. \cite{Kataev:1996vu}.

\section{Available perturbative corrections}

The status of the  $O(\alpha_s)$ perturbative 
QCD corrections to $I_{GSR}$ was summarised in Ref. \cite{Hinchliffe:1996hc}.
Following this review, we will extend its presentation to the 
order  $\alpha_s^2$ level.

It should be stressed that the renormalisation group equation for 
$I_{GSR}$ contains the anomalous dimension term:
\begin{equation}
\bigg[\mu\frac{\partial}{\partial\mu}+\beta(A_s)\frac{\partial}
{\partial A_s}-\gamma_{I_{GSR}}^{n=1}(A_s)\bigg]C_{I_{GSR}}(A_s)=0
\end{equation}     
where $A_s=\alpha_s/(4\pi)$ and
\begin{equation}
\mu\frac{\partial A_s}{\partial \mu}=-2\sum_{i\geq 0}\beta_i A_s^{i+2}
\end{equation}
The first two scheme-independent  coefficients in Eq. (6) are well-known:
\begin{eqnarray}
\beta_0&=&\bigg(\frac{11}{3}C_A-\frac{2}{3}n_f\bigg)  
        = 11 -0.666667n_f \\  
\beta_1&=&\bigg(\frac{34}{3}C_A^2-2C_Fn_f-\frac{10}{3}C_An_f\bigg)
       = 102-12.6667n_f   
\end{eqnarray}
where $C_F=4/3$, $C_A=3$ and $n_f$ is the number of active flavours.

The corresponding  anomalous dimension function has the canonical  expansion
\begin{equation}
\gamma_{I_{GSR}}^{n=1}=\sum_{i\geq 0} \gamma_i^{n=1}A_s^{i+1}~~~.
\end{equation}
However,  like in the case of the first moments of  
SFs of  $\nu N$ DIS, the first 
coefficient of  the NS anomalous dimension function  
of the first moment 
$\gamma_0^{n=1}$ is identically equal to zero. The difference is starting 
to manifest itself from the two-loop level, where in order to get 
the corresponding result in case of anomalous 
dimension for $I_{GSR}$  it is necessary to make analytical continuation   
and to
use the so-called (+) prescription (see e.g. Ref. \cite{Yndurain}). 
In the case of $\gamma_1^{n=1}$ this was done in Ref. \cite{Ross:1978xk}
and Ref. \cite{Curci:1980uw} and result in the following analytical 
expression
\begin{equation}
\gamma_1^{n=1}=-4(C_F^2-C_FC_A/2)[13+8\zeta(3)-2\pi^2]= +2.55755
\end{equation} 
where  the numerical value of  $\zeta(3)$=1.2020569 was 
taken into account.
The perturbative corrections to  $I_{GSR}$ can be obtained 
from the solution of the renormalisation group equation of Eq. (5):
\begin{equation}
I_{GSR}(A_s)=AD(A_s)\times C(A_s)
\end{equation}
where the anomalous dimension term is defined as 
\begin{equation}
\label{AD}
AD(A_s)=exp\bigg[-\int_\delta^{A_s(Q^2)}\frac{\gamma^{n=1}_{I_{GSR}}(x)}{\beta(x)}dx
\bigg]~~~.
\end{equation}
Since the first coefficient of $\gamma^{n=1}_{I_{GSR}}$ is identically 
zero ( namely $\gamma_0^{n=1}$=0), there is no singularity in $AD(A_s)$ 
and we can put in Eq. (\ref{AD}) the lower bound of integration $\delta=0$.
In this case we obtain the following  
expression for the expansion 
of $AD(A_s)$ up to $O(\alpha_s^2)$-corrections: 
\begin{equation}
\label{EAD}
AD(A_s(Q^2))=1+\frac{1}{2}\frac{\gamma_1^{n=1}}{\beta_0}A_s(Q^2)+
\frac{1}{4}\bigg(\frac{1}{2}\frac{(\gamma_1^{n=1})^2}{\beta_0^2}-
\frac{\gamma_1^{n=1}\beta_1}{\beta_0^2}+\frac{\gamma_2^{n=1}}{\beta_0}\bigg)
A_s^2(Q^2)~~~.
\end{equation}
The only unknown terms here is the third coefficient $\gamma_2^{n=1}$ 
of the  anomalous dimension function $\gamma_{I_{GSR}}^{n=1}(A_s)$,
which in general is scheme-dependent.

In the cases of $n_f=3$ and $n_f=4$ the  numerical versions of Eq. (\ref{EAD})
 read
\begin{eqnarray}
AD(\alpha_s)_{n_f=3}&=&1+0.0355\bigg(\frac{\alpha_s}{\pi}\bigg)
+\bigg(-0.0392+\frac{\gamma_2^{n=1}}{64\beta_0}\bigg)\bigg(\frac{\alpha_s}{\pi}\bigg)^2 \\ 
AD(\alpha_s)_{n_f=4}&=&1+0.0384\bigg(\frac{\alpha_s}{\pi}\bigg)
+\bigg(-0.0415+\frac{\gamma_2^{n=1}}{64\beta_0}\bigg)\bigg(\frac{\alpha_s}{\pi}\bigg)^2 
\end{eqnarray} 
where the scheme-dependent    
expression for $\gamma_2^{n=1}$ is still unknown. Its  
value will be fixed in the next section using the results of calculations 
in the $\overline {\rm MS}$-scheme.

Few words should be added here on the perturbative theory expansion 
of $C(A_s)$. From the general grounds it should have the following form:
\begin{equation}
C(A_s)=\frac{1}{3}\bigg[1+C_1^{n=1}A_s(Q^2)+C_2^{n=1}A_s^2(Q^2)\bigg]~~~.
\end{equation}
As was found in Ref. \cite{Bardeen:1978yd} its first coefficient is zero, 
namely $C_1^{n=1}=0$. However, as will be shown in the next section 
the non-zero perturbative theory contribution is appearing at 
the two-loop level.

\section{Calculations and estimates of the $\alpha_s^2$ contributions}

We will start 
from the calculations 
of perturbative contribution to the 
coefficient function $C(A_s)$ at the $\alpha_s^2$-level. 
It can be obtained after applying (+) prescription to the results of 
Ref. \cite{vanNeerven:1991nn}.
Indeed, the order $\alpha_s^2$ correction to the coefficient function 
of $I_{GSR}$ is defined by taking the first moment from the 
sum  
\begin{equation}
\label{sum}
C_2^{n=1} =\int_0^1\bigg[ C_2^{(2),(-)}(x,1)+C_2^{(2),(+)}(x,1)\bigg]dx~~~~,
\end{equation}
 where the expressions for the 
functions  
$C_2^{(2),(-)}(x,1)$ and    $C_2^{(2),(+)}(x,1)$ were calculated in
Ref. \cite{vanNeerven:1991nn} and confirmed with the help of another 
technique in Ref.\cite{Moch:1999eb}.  
Integrating Eq. (\ref{sum}) 
numerically with 
arbitrary Casimir operators $C_A$ and $C_F$, we obtain the following 
$n_f$-independent and scheme-independent  result
\begin{eqnarray}
C(A_s)&=&\frac{1}{3}\bigg[1-0\bigg(\frac{\alpha_s}{\pi}\bigg)+
(3.695C_F^2-1.847C_FC_A)\bigg(\frac{\alpha_s}{\pi}\bigg)^2\bigg] \\ \nonumber
&=&\frac{1}{3}\bigg[1-0.821
 \bigg(\frac{\alpha_s}{\pi}\bigg)^2\bigg]~~~~.
\end{eqnarray}
Combining now Eq. (14) and Eq. (15)  with Eq. (18) we find
the following expressions for $I_{GSR}$:
\begin{eqnarray} 
I_{GSR}(Q^2)_{n_f=3}&=&\frac{1}{3}
\bigg[1+0.0355\bigg(\frac{\alpha_s}{\pi}\bigg)
+\bigg(-0.862+ \frac{\gamma_2^{n=1}}{64\beta_0}\bigg)\bigg
(\frac{\alpha_s}{\pi}\bigg)^2\bigg] \\ 
I_{GSR}(Q^2)_{n_f=4}&=&\frac{1}{3}
\bigg[1+0.0384 \bigg(\frac{\alpha_s}{\pi}\bigg)
+\bigg(-0.809 + \frac{\gamma_2^{n=1}}{64\beta_0}\bigg)\bigg
(\frac{\alpha_s}{\pi}\bigg)^2\bigg]
\end{eqnarray}
where $\alpha_s=\alpha_s(Q^2)$ is the NLO expression for $\overline {\rm MS}$ 
coupling constant.

In order to get the feeling what might be the contribution 
of the terms proportional to $\gamma_2^{n=1}$ we will avoid 
extrapolation 
procedure of the  values of $\gamma_{2}^{n}$ used in Ref.\cite{Kataev:1996vu}, 
calculated 
analytically for even n=2,4,..,14 in the works of Ref. \cite{Larin:1993vu}.
Indeed, performing extrapolation from the even values of $n$ 
for the NLO terms  $\gamma_{1}^{n}$ of the corresponding anomalous 
dimension function,  we are obtaining the following estimate  
$\gamma_1^{n=1}=28.23$, which is 10 times larger that the real value 
given in Eq. (10). Therefore, the used in Ref.\cite{Kataev:1996vu}
extrapolation  procedure is considerably overestimating
the value of the coefficient  $\gamma_1^{n=1}$. The similar situation can 
occur in the case of using extrapolation procedure for fixing the value 
of $\gamma_2^{n=1}$. Indeed, following the ideas of Ref.\cite{Kataev:1996vu}
we get from extrapolation of the known even values for $\gamma_2^{n}$ 
the following estimates:  $\gamma_2^{n=1}\approx361$ for $n_f=3$ and 
 $\gamma_2^{n=1}\approx283$ for $n_f=4$, which to our point of view might  
be unrealistically large. 

Keeping in mind that only direct calculation of $\gamma_2^{n=1}$
can give the real numerical value of this term, we nevertheless
are proposing the following way of fixing uncalculated contribution 
to $\gamma_{I_{GSR}}^{n=1}$ function.
We noticed the following  numerical pattern of the behaviour of 
anomalous dimension 
function for $n\geq 2$: 
 $\gamma_1^{n}/\gamma_2^{n}\sim 0.12$ for $n_f$=4 
(see Ref.\cite{Kataev:1996vu} and Ref.\cite{Kataev:2001kk} especially).
We have checked that for $n_f$=3 the similar relation is  
$\gamma_1^{n}/\gamma_2^{n}\sim 0.09$ for $n\geq 2$.
Hoping that these relations   are  also valid in case of $n\geq 1$ , we
estimate the values for $\gamma_2^{n=1}=(1/0.12) \gamma_1^{n=1}
\approx 21.3$
in the case of $n_f=4$ and  $\gamma_2^{n=1}=(1/0.09) \gamma_1^{n=1}
\approx 28.4$ in the case of $n_f=3$. 
Substituting them into Eq.(19) and Eq.(20) we get: 
 \begin{eqnarray} 
I_{GSR}(Q^2)_{n_f=3}&=&\frac{1}{3}\bigg[1+0.0355\bigg(\frac{\alpha_s}{\pi}\bigg)
-0.811 \bigg
(\frac{\alpha_s}{\pi}\bigg)^2\bigg] \\ 
I_{GSR}(Q^2)_{n_f=4}&=&\frac{1}{3}\bigg[1+0.0384 \bigg(\frac{\alpha_s}{\pi}\bigg)
-0.822
\bigg(\frac{\alpha_s}{\pi}\bigg)^2\bigg]~
~~.
\end{eqnarray}
Taking now $\alpha_s(Q^2)\approx 0.35$
we arrive to the following numerical versions of Eq.(21) and Eq.(22):
\begin{eqnarray}
I_{GSR}(Q^2)_{n_f=3}&=&\frac{1}{3}\bigg[1+0.0039-0.0101\bigg]=0.3313 \\ 
I_{GSR}(Q^2)_{n_f=4}&=&\frac{1}{3}\bigg[1+0.0042 - 0.0102\bigg]=0.3313
\end{eqnarray}
Therefore in presented expression for  the the order
$\alpha_s^2$ correction to the Gottfried sum rule 
is larger then the order $\alpha_s$-term.

Theoretical errors to the presented third terms 
in Eqs.(21)-(34)  are  coming from the errors of 
$\gamma_2^{n=1}$ terms in Eqs. (19), (20), which is impossible to estimate 
without their direct theoretical calculations. In any case these terms 
are damped by huge numbers $(64\beta_0)$ and it is unlikely that 
the direct calculations of  $\gamma_2^{n=1}$ terms will change 
the results of Eqs.(23), (24) substantially. 
One can check this conclusion using the overestimated to our point 
of view results of application of the extrapolation procedure.
Moreover, the main contributions to the 
$\alpha_s^2$-term in Eqs. (21)-(24) come from the calculated 
by us $\alpha_s^2$ term
of the coefficient function of the Gottfried sum rule.

\section{Comments on violation of the Gottfried sum rule} 
In the previous section we found that 
order $\alpha_s^2$  perturbative QCD  
corrections to the Gottfried sum rule are really small 
and can not describe violation of the theoretical prediction 
from its NMC experimental value. 
This, in turn, lead to the  necessity   of introduction of the effect 
of flavour asymmetry
of antiquark distributions in the  nucleon \cite{Arneodo:1994ia}, namely 
\begin{equation}
\int_0^1dx \bigg[\overline{d}(x,4~{\rm GeV}^2)-
\overline{u}(x,4~{\rm GeV}^2)\bigg]_{NMC}= 0.147\pm 0.039~~~.
\end{equation}
This phenomenological result is important for fixing the  corresponding 
$\overline{d}/\overline{u}$ ratio in different sets of  parton distribution
functions, which are relevant to the  LHC physics 
(for a review see e.g. Ref. \cite{Catani:2000jh} ).
On the other side the consideration of available E866 data for the Drell-Yan 
production in proton-proton and proton-deuteron scattering has confirmed 
the effects of flavour asymmetry.
Indeed the analysis of Ref. \cite{Towell:2001nh}
gave the following 
number 
\begin{equation}
\label{E866}
\int_{0.015}^{0.35}dx \bigg[\overline{d}(x,54~{\rm GeV}^2)-
\overline{u}(x,54~{\rm GeV}^2)\bigg]_{E866}= 0.0803 \pm 0.011~~~.
\end{equation}
It was also  noted in Ref. \cite{Towell:2001nh} 
that it is unlikely to receive 
additional contribution to Eq. (\ref{E866})from the region above $x=0.35$, 
since the sea is rather small in this region. However, the contribution 
to this whole integral from the unmeasured region $x \leq 0.015$ is 
missed.  
The attempt to fix  it was made in Ref. \cite{Szczurek:1999wp}
using the extrapolation to small $x$ region. As the result 
the authors of Ref. \cite{Szczurek:1999wp}
suggested the manifestation of substantial contribution of twist-4 
$1/Q^2$-effects in Eq. (1). Note,  that final E866 result 
is 
\begin{equation}
\int_0^1dx \bigg[\overline{d}(x,54~{\rm GeV}^2)-
\overline{u}(x,54~{\rm GeV}^2)\bigg]_{E866}= 0.118\pm 0.012~~~.
\end{equation}
is closer to NMC result, than obtained in Ref.\cite{Szczurek:1999wp}
extrapolated value, namely 
\begin{equation}
\int_0^1dx \bigg[\overline{d}(x,50~{\rm GeV}^2)-
\overline{u}(x,50~{\rm GeV}^2)\bigg]_{Ref. \cite{Szczurek:1999wp}}
= 0.09\pm 0.02~~~.
\end{equation}
Therefore, in order to understand the status 
of their  conclusion on the possibility of existence of substantial 
contribution 
of the $1/Q^2$-corrections to the Gottfried sum rule 
it is necessary to be more careful in performing extrapolations 
to low $x$-region. It is  
highly desirable to estimate the effects 
of higher-twist contributions to the Gottfried sum rule using any 
concrete model.
However, one should keep in mind that there are also some other 
explanations of the observed deviation from the canonical value $1/3$ 
for the Gottfried sum rule (see Ref. 
\cite{Karliner:2002pk} and Ref. \cite{Kumano:1997cy} for the review 
of other works 
on the subject).

\section{Conclusions}
In summary:
we found  non-zero  $O(\alpha_s^2)$ 
perturbative QCD contributions to the\\
coefficient function of the  Gottfried sum rule. 
We  also estimated 
the effect due to non-zero value of the three-loop contribution to 
NS anomalous dimension function for $n=1$ moment, which as we think 
is rather small.
More detailed result 
can be obtained after completing  the analytical calculations of the 
three-loop corrections 
to the NS kernel of DGLAP equation. This work is  
now in progress (see e.g. Ref.  \cite{Moch:2002sn}). 
In any case the value of the  $\alpha_s^2$-correction  
is   dominated 
by the calculated in this Letter contribution to the coefficient function, 
which is negative, but  also small.  Therefore, the existing NMC observation  
of the flavour asymmetry between $\overline{d}$ and $\overline{u}$ antiquarks 
is surviving. We hope that the possible future new HERA data might be 
useful for more detailed measurement of the Gottfried sum rule and for 
further studies of the effect of flavour asymmetry in the 
$\overline{d}/\overline{u}$ ratio.

\section{Acknowledgements}
We are grateful to J. Blumlein and W. van Neerven for stimulating us 
to reconsider the effects of contributions of the order $O(\alpha_s^2)$ 
corrections to the Gottfried sum rule. We would like to thank 
S.I. Alekhin  for discussions. 
The work of ALK was done within 
scientific program of RFBR Grants N 02-01-00601,   03-02-17047 and  03-02-17177,
The work of GP was supported by Galician research funds (PGIDT00 PX20615PR) 
and Spanish CICYT (FPA 2002-01161).

\end{document}